\documentclass[namedreferences,hyperref,optionalrh]{spr-sola}
\usepackage{graphicx}        
\usepackage{amssymb}  
\usepackage{amsmath}
\usepackage{color}           
\usepackage{breakurl}                         

\chardef\us=`\_

\begin{document}
\begin{frontmatter}

\title{Correlation of Coronal Hole Area Indices and Solar Wind Speed}

\author[addressref={aff1,aff2},email={egor.illarionov@math.msu.ru}]{\inits{E.A.}\fnm{Egor}~\lnm{Illarionov}\orcid{0000-0002-2858-9625}}

\author[addressref=aff3,email={}]{\inits{A.G.}\fnm{Andrey}~\lnm{Tlatov}\orcid{0000-0002-6286-3544}}

\author[addressref=aff3,email={}]{\inits{I.A.}\fnm{Ivan}~\lnm{Berezin}\orcid{0000-0002-6876-6761}}

\author[addressref=aff3,email={}]{\inits{N.N.}\fnm{Nadezhda}~\lnm{Skorbezh}\orcid{0000-0003-0465-7133}}

\address[id=aff1]{Moscow State University, Moscow, Russia}
\address[id=aff2]{Institute of Continuous Media Mechanics, Perm, Russia}
\address[id=aff3]{Kislovodsk Mountain Astronomical Station of the Pulkovo Observatory, Kislovodsk, Russia}

\runningauthor{Illarionov et al.}
\runningtitle{Coronal hole index}

\begin{abstract}
Coronal holes (CHs) are widely considered as the main sources of high-speed solar wind streams. We validate this thesis comparing the smoothed time series of solar wind speed measured by Advanced
Composition Explorer (ACE) and various indices of CH areas constructed from the CH catalog compiled at the Kislovodsk Mountain Astronomical Station for the period 2010--2025. The main result is that we find specific indices of CH areas that give a strong correlation with smoothed solar wind speed variations. As an example, 1-year averaged areas of CHs located within 30 degrees of the solar equator yield a correlation of 0.9 with 1-year averaged solar wind speed. This strong correlation is a feature of the particular CH catalog, and considering an alternative CH catalog obtained using the Spatial Possibilistic Clustering Algorithm (SPoCA) from the Heliophysics Event Knowledgebase (HEK), the same index provides a correlation of only 0.3. Although the fact that the correlation significantly depends on the catalog requires a separate discussion, we conclude that if some of the catalogs can be used to construct a reliable indicator of solar wind speed variations, then this methodology should be maintained further. Additionally, we present time-latitude diagrams of rolling correlation between CHs areas and solar wind speed, which, in our opinion, can be used to reveal source CHs for high-speed solar wind streams.
\end{abstract}
\end{frontmatter}

\section{Introduction}

Coronal holes (CHs) are large-scale long-lived structures on the Sun that arise in unipolar regions with an open configuration of magnetic field lines and are observed in the extreme ultraviolet and X-ray wavelength as regions of low brightness 
\citep{Tousey, Cranmer, Harvey}.
It has been shown \citep[e.g.,][]{Krieger, Sheeley} that coronal holes are the source of high-speed solar wind streams (HSS), which, along with coronal mass ejections, lead to the development of geomagnetic disturbances.
Coronal holes can exist for several solar rotation periods, which leads to the formation of recurrent solar wind streams.

Further investigation of the connection between CHs and HSS usually starts with association of peaks in solar wind speed with particular CHs and correlation of parameters of the CHs with the solar wind speed. This approach is specifically important for operational space weather prediction. But due to the difficulties in determination of the source CH for each solar wind speed peak, it is usually validated on a small subset of CHs \citep{Nolte, vrvsnak2007coronal, Abramenko, Karachik, Rotter, 2018Hofmeister, 2018Heinemann, 2020Heinemann, Evangelia}. For large time intervals (e.g., several solar cycles) the process becomes complicated. Moreover, on large time scales instantaneous variations of solar wind speed become less important, and one can apply time averaging. Using time averaging, one can also avoid the problem of a delay between CH observation and arrival of HSS as well as the contribution of other sources to formation of HSS. We will focus on time-averaging windows of about 1 year length that is common for another index of solar activity, the sunspot number.

Although the process of time averaging of solar wind speed data is quite straightforward, it is not clear what index derived from CHs should be used for comparison. The goal of this study is to investigate various indices and various time scales and to understand to what extent CH data can explain (approximate) smoothed variations of solar wind speed.

It is also expected that the potential CH index can depend on the catalog of CHs due to variety of methodologies of CH identification 
\citep[see][for review and comparison]{Reiss_2021,Reiss_2024}.
The problem of CH identification is complicated by the presence of dark regions of quiet Sun, filaments, and transient dimmings, which have the same intensity range as CHs \citep{Caplan}. The variety of approaches to resolve these ambiguities leads to a variety of results and,
as was shown in \cite{Reiss_2021} and \cite{Reiss_2024}, the difference in CHs detected with different algorithms can be of two or more times.

We will use the CH dataset prepared at the Kislovodsk Mountain Astronomical Station and the SPoCA-HEK catalog \citep{SpoCA2014} for comparison. Both catalogs are based on solar disk observations made by Solar Dynamics Observatory (SDO)
Atmospheric Imaging Assembly (AIA) telescope \citep{Lemen2012} in 193 Angstrom wavelength for the period 2010--2025. Various indices of CH areas will be compared with the solar wind speed data measured by Advanced
Composition Explorer \citep[ACE,][]{Stone1998} located at the L1 libration point.

\section{Coronal Hole Data}

For construction of CH indices we will use only basic parameters of the CHs such as the latitude of the center of CH and area. The primary catalog for this research will be the catalog prepared at Kislovodsk Mountain Astronomical Station (KMAS). 
The CH identification methodology at KMAS is maintained over a long period of time and is based on SDO/AIA extreme ultraviolet solar disk images. 

Specifically, SDO/AIA 193~Angstrom FITS images are taken for every day between 2010 and 2025 at 10~UT. 
The images are downsized from $4096\times 4096$ pixels to a size of $1024\times 1024$ and normalized. 
For normalization, the point of the maximum of the intensity distribution in the solar disk, where 
$R<0.8R_{\odot}$, denoted $I_{\rm {disk}}$, and the point of the
maximum of the intensity distribution in the solar corona at a distance of $1.0R_{\odot}<R<1.2R_{\odot}$, denoted $I_{\rm {cor}}$, are calculated. The new intensity for a pixel with intensity $I$ is calculated as 
$I_{\rm {new}}=10\log_{10}
\left(\max(0.1, 100I_{\rm disk}(I -I_{\rm {cor}})/(I_{\rm {disk}}-I_{\rm {cor}})\right)$ for all image pixels.

For CH identification, the iterative region growing procedure is applied. 
The idea is that starting from the pixel with the lowest intensity (region),
surrounding pixels with intensities below some local threshold (updated at each step)
are attached to the CH region. In the initial steps of this process, the area of the region grows smoothly,
but at some iteration, when the threshold becomes high enough, the area increases sharply,
and this is the trigger to stop the process. Once the process is finished, the
pixels attributed to the CHs are excluded from the image and the process is repeated
for the remaining pixels. Finally, CHs with areas below 2\,000 millionth of the
solar hemisphere (MSH) corresponding to about $6.1\times10^9$ $\rm{km}^2$ are filtered out.
Finer details of the process can be found in \cite{Tavastsherna}.
The obtained catalog of CHs is presented and daily updated at the website \url{https://www.observethesun.com}.

For comparison, the SPoCA-HEK catalog \citep{SpoCA2014} was obtained from the
Heliophysics Event Knowledgebase (HEK, \url{https://www.lmsal.com/hek}) for each day in 2010--2025. In this catalog, we removed CHs with areas below 2\,000 MSH as well as it is done in the KMAS catalog.

\begin{figure}
\centering
\includegraphics[width=0.97\columnwidth]{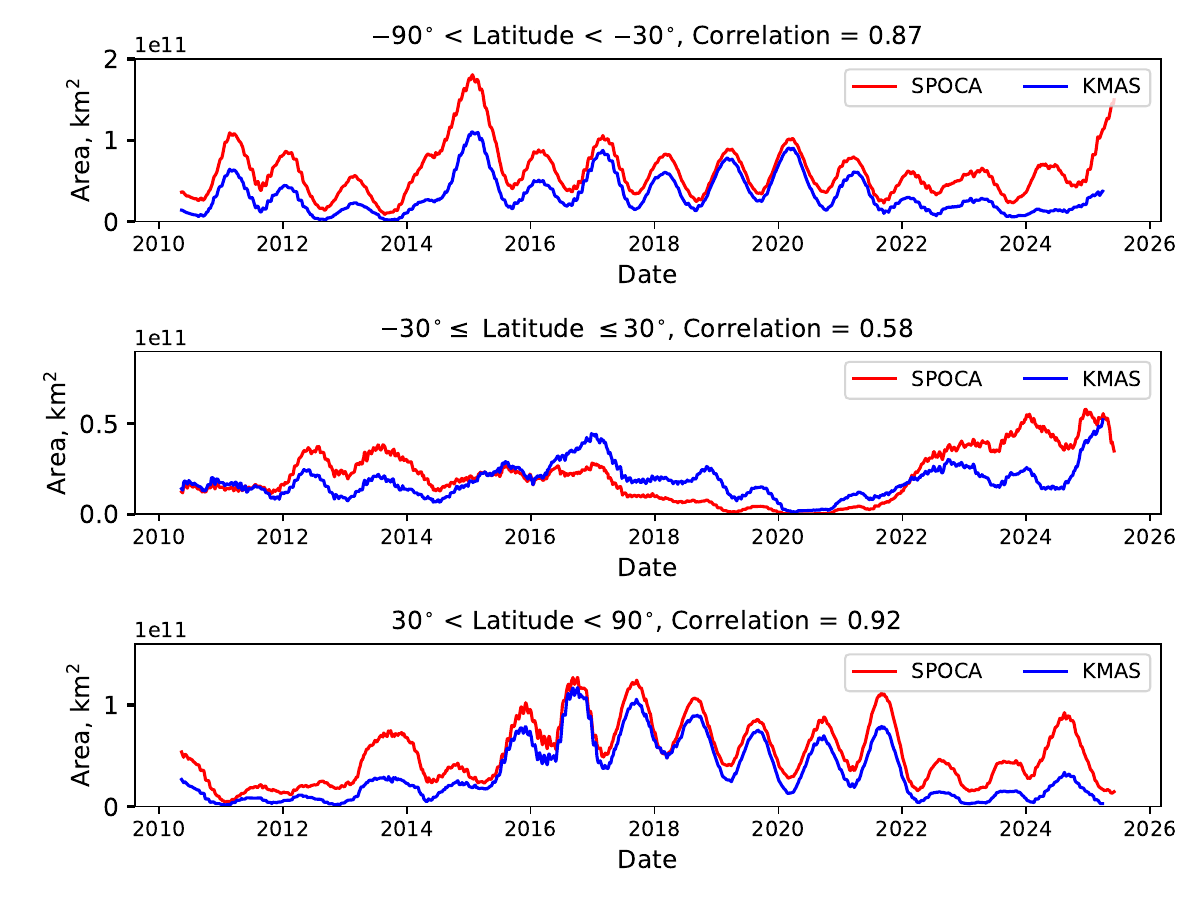}
\caption{Total daily area of coronal holes averaged over sliding windows of size 180 days. The red line is for CHs presented in the
SPoCA-HEK catalog, the blue line shows CHs from the catalog of the Kislovodsk Mountain Astronomical Station.
The title of each plot indicates the latitude interval used to select CHs and the correlation coefficient between both time series. Areas are scaled to $10^{11}$ $\rm{km}^2$.}
\label{chs}
\end{figure}

Figure~\ref{chs} shows the total area of CH in different latitudinal intervals (north and south polar regions and low-latitude region) averaged using a 180-day (1/2-year) sliding window.
The agreement between both catalogs is better for polar CHs (correlation
is about 0.9) and lower for low-latitude CHs (correlation is about 0.6). 
We also note that the total area of polar CHs is systematically larger in the SPoCA-HEK catalog, while for low-latitude CHs the situation changes with time.

Figure~\ref{chs} also demonstrates our motivation for selecting the SPoCA-HEK catalog for comparison. This catalog is quite close in terms of the total CHs to the KMAS catalog, but despite this, the difference in correlation with solar wind data will be significant as we demonstrate below.

\section{Solar Wind Data}

Daily solar wind speed data measured by ACE are obtained from the OMNI database
\href{https://omniweb.gsfc.nasa.gov}{https://omniweb.gsfc.nasa.gov}.
In this study we are interested in a correlation of high-speed solar wind streams (HSS) with CH parameters.
However, HSS may be also connected with interplanetary coronal mass ejections (ICMEs).
It looks reasonable to exclude such events, but as we show below, it has only a minor effect
when averaging data over wide time intervals.

To investigate the impact of ICME, we use the catalog of near-Earth ICMEs prepared by \cite{Richardson} and obtained
at \url{https://srl.caltech.edu/ACE/ASC/DATA/level3/icmetable2.htm}. 
We select ICMEs with a maximum flow speed (given by $V_{max}$ parameter) greater than 400 km/s, and drop all ACE data in the period one day before and three days after the start date of the ICMEs (five days in total). 
This operation produces missing values in ACE time series data.
To compare the original ACE data and the ACE data without ICMEs, we averaged both time series using a 180-day sliding window.
Note that when averaging, missing values are not filled with zeros, but are ignored both in the numerator or denominator of the averaging formula. The resulting values are assigned to the center points of the sliding windows.

\begin{figure}
\centering
\includegraphics[width=0.9\columnwidth]{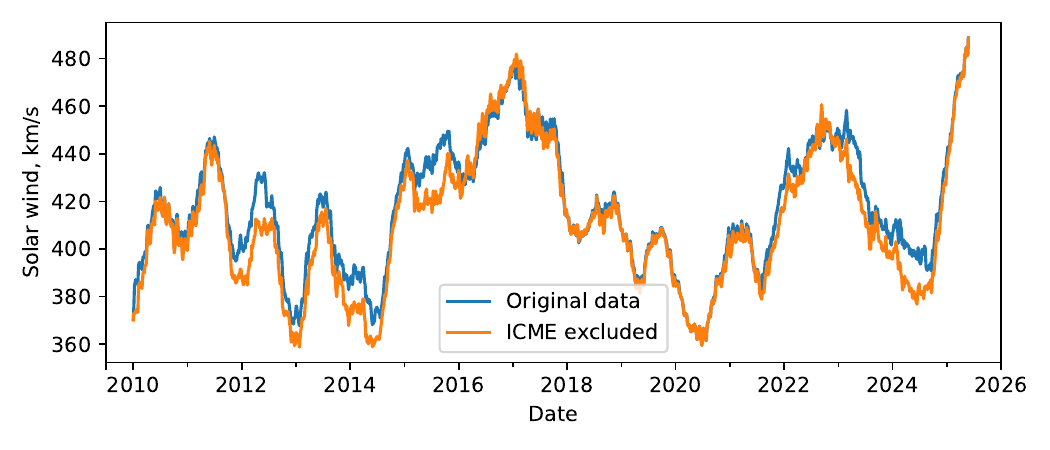}
\caption{Solar wind speed averaged over 180-days sliding windows. The blue line shows averaging of the
original ACE data, the orange line shows averaging of the ACE data with excluded values during ICMEs.}
\label{solar_wind}
\end{figure}

Figure~\ref{solar_wind} shows a comparison of the original ACE data and the ACE data with excluded ICMEs.
We note that the difference is quite moderate, and for the sake of simplicity we will use the original
ACE data without any filtering in the next sections of the paper.

\section{Results: Correlation with Solar Wind Speed}

Although CH parameters and CH boundaries enable construction of complex morphological indices, we consider the most trivial one, the total daily area of CHs within certain latitudinal bounds, averaged over various time windows.

We consider three latitudinal zones: (i) $|\theta| \le 30^{\circ}$, (ii) $|\theta| \le 50^{\circ}$, and (iii) $|\theta| \le 90^{\circ}$. The boundary $50^{\circ}$ is motivated by its use as a boundary separating polar CHs \citep[see, e.g.,]{HessWebber2014}. Then we compute the total CH area for each day for each latitudinal zone (if there are no CHs for a particular day, we set the total area to zero).
Finally, we average the daily values using sliding windows of size (i) 1/2 year, (ii) 1 year, and (iii) 2 years. In total, we obtain nine variants of CH indices and correlate them with solar wind speed averaged over the same time intervals.

\begin{figure}
\centering
\includegraphics[width=0.975\columnwidth]{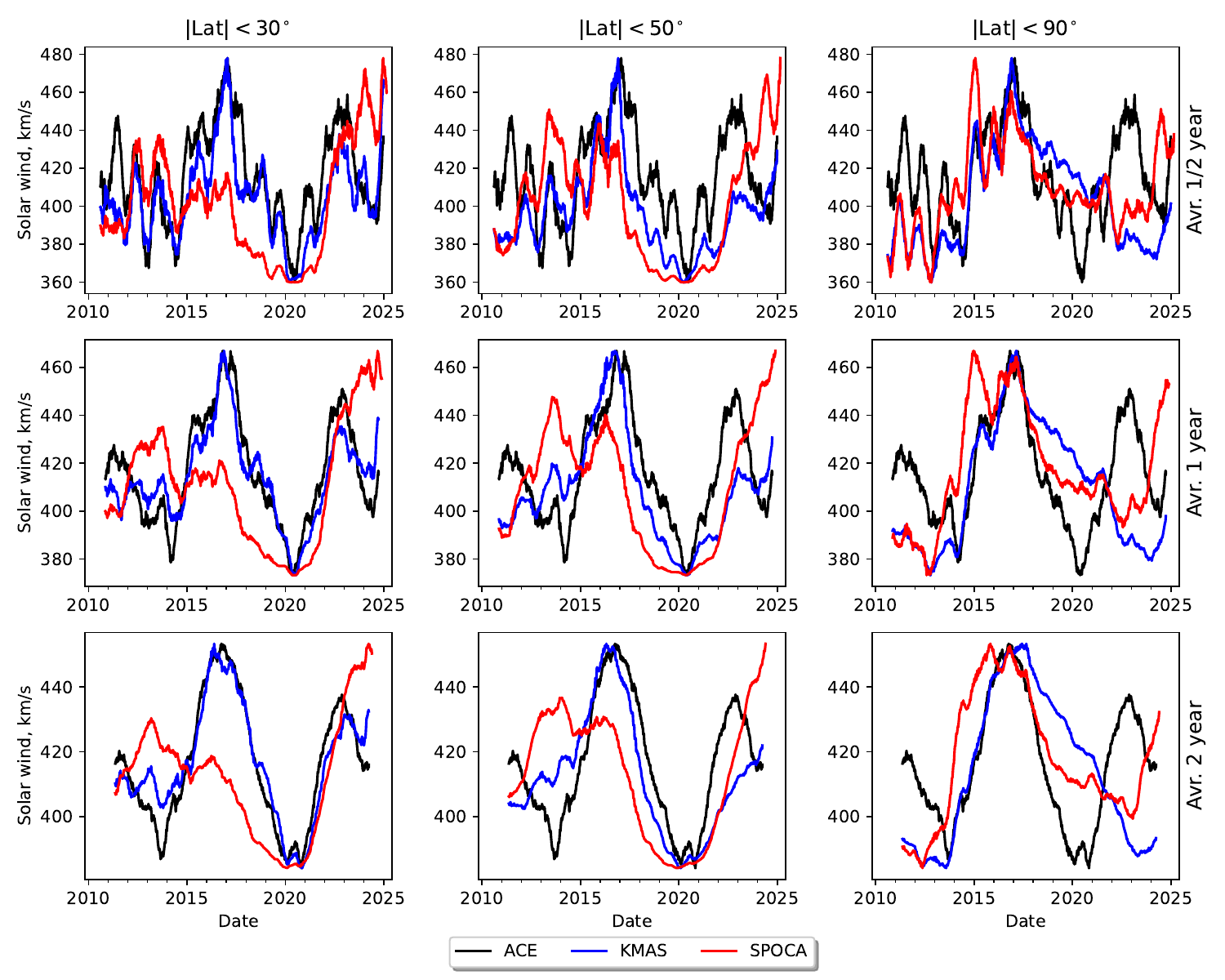}
\caption{Various indices of coronal hole areas compared with the solar wind speed. The blue lines show indices constructed from  the KMAS catalog, red lines are for indices constructed from SPoCA-HEK catalog. The black lines show solar wind speed measured by ACE. In the first row, indices and solar wind speed data are averaged over 1/2-year sliding windows, in the second row -- over 1-year sliding windows, in the last row -- over 2-years sliding windows. The first column shows the total areas of CH within $30^{\circ}$ of the solar equator, in the second column -- within $50^{\circ}$, and from the whole solar disk in the third column.}
\label{corr_plots}
\end{figure}

\begin{figure}
\centering
\includegraphics[width=0.85\columnwidth]{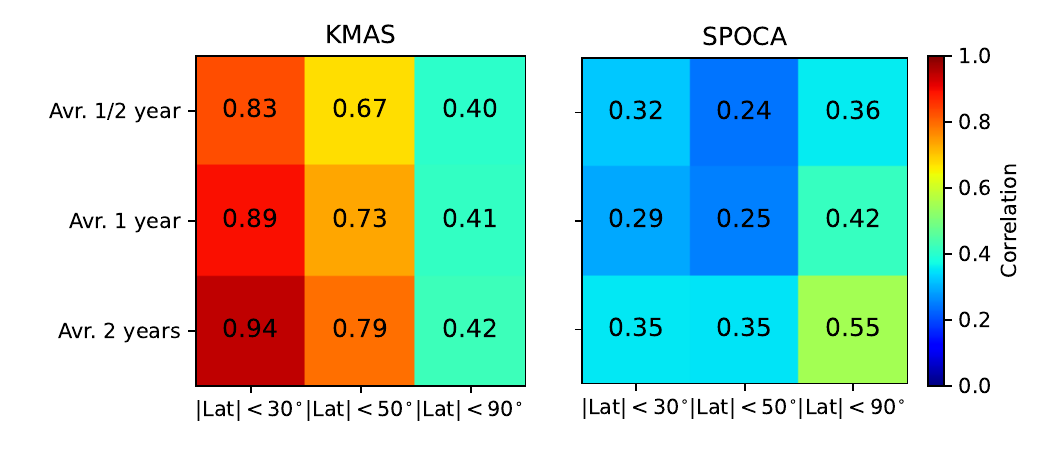}
\caption{Correlation coefficients between CH indices and solar wind speed. Left panel is for KMAS catalog of CHs, right panel in for SPoCA-HEK catalog. Rows and columns in each panel correspond to rows and columns in Figure~\ref{corr_plots}.}
\label{corr_tables}
\end{figure}

Figure~\ref{corr_plots} shows comparison of the nine constructed indices with solar wind speed variations. In this figure, the CH indices representing CH area, measured in $\rm{km}^2$, are linearly scaled to fit the range of solar wind speed, measured in km/s.
The correlation coefficients for each index are shown in Figure~\ref{corr_tables}.

We observe in Figures~\ref{corr_plots} and~\ref{corr_tables} that for KMAS catalog reduction of the latitudinal zone from $90^{\circ}$ (whole solar disk) to lower latitudes ($30^{\circ}$) uniformly increases the correlation for all time-averaging windows. Increasing of the time-averaging window leads to increase of the correlation coefficients in all latitudinal domains. In particular, a correlation of 0.9 is obtained for the index, constructed from CHs within $30^{\circ}$ of the solar equator and averaged over 1-year sliding window. Averaging over 2-years sliding windows yields a correlation of 0.94. The explicit formula for the CH index, including scaling coefficients, reads:
\begin{equation}
    I_{CH} = (2.7/10^{9})\left[1/(\rm{km}\times s)\right]\times Area\left[\rm{km}^2\right] + 365\left[\rm{km}/s\right] \,,
\end{equation}
where $Area$ is the result of 1-year (or 2-year) averaging of total daily areas of CHs whose central latitude is within $30^{\circ}$ of the solar equator. The value $I_{CH}$ is attributed to the central point of the time-averaging window.

The comparison with the SPoCA-HEK catalog demonstrates that the high correlation for low latitudes and $\ge 1$ year averaging is not a trivial property of exclusion of the polar CHs and large averaging window, but a feature of the catalog. Indeed, in the SPoCA catalog, the correlation does not exceed 0.55 for all considered combinations of latitudinal boundary and time averaging window. Moreover, we observe that the maximal correlation (0.55) is obtained for the CHs from the whole disk, i.e., including polar CHs, while for low-latitudinal CHs the correlation is about 0.3 only. Note that both catalogs yield quite similar moderate correlation ($\approx 0.4-0.5$) for the CHs from the whole disk.

\begin{figure}
\centering
\includegraphics[width=0.95\columnwidth]{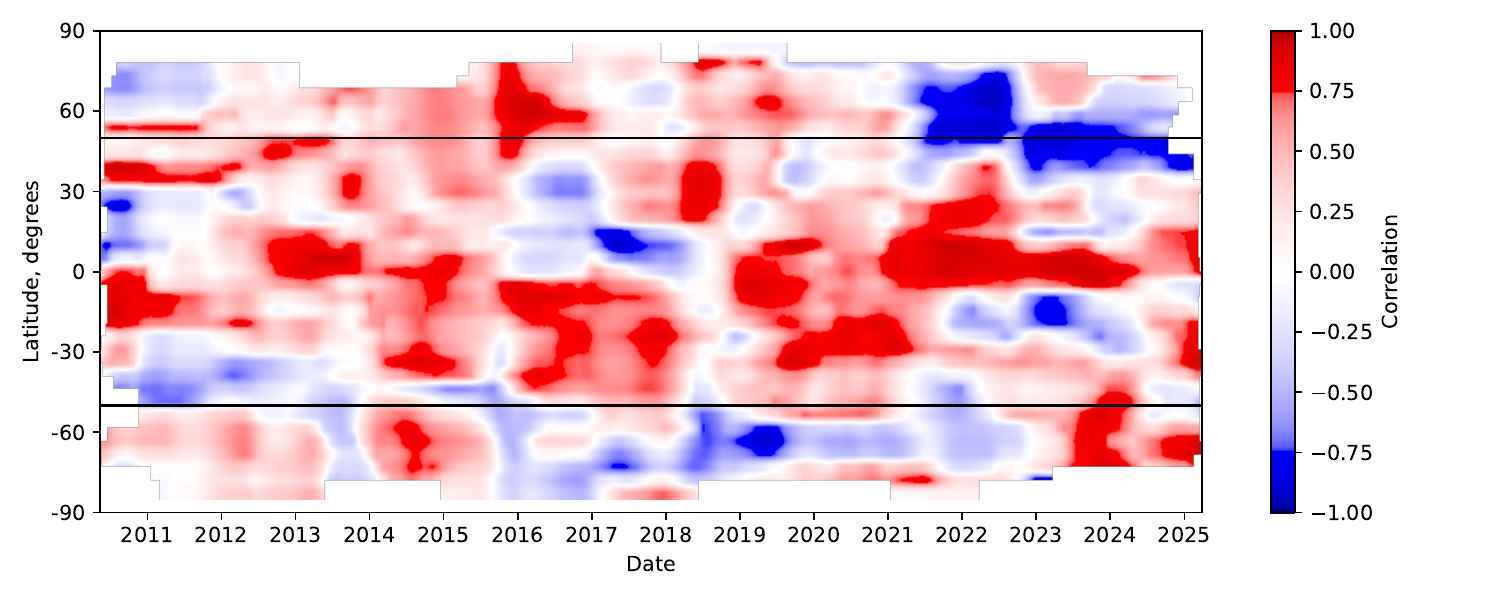}
\caption{Time-latitude diagram of rolling correlation coefficients between area of CHs around given latitude and around given moment of time and solar wind speed. Data are averaged using 1/2-year sliding windows; rolling correlation coefficients are computed using sliding windows of size 2 years. Latitudinal bins have a size of $10^{\circ}$. CHs are obtained from the KMAS catalog. Black horizontal lines indicate $\pm 50^{\circ}$ latitudes.}
\label{corr_map}
\end{figure}

The strong correlation obtained using the KMAS catalog motivates us to investigate it in more detail. We consider the following question: for a given moment of time, CHs from what latitudes are most correlated with solar wind speed. To address this question, we construct a time-latitude diagram (Figure~\ref{corr_map}), where each value is a correlation coefficient between areas of CHs around this latitude and around this moment of time and solar wind speed.
More detailed, the algorithm for construction of time-latitude diagram (Figure~\ref{corr_map}) is as follows:
\begin{itemize}
    \item latitudes are split into overlapping bins ($90^{\circ}$, $80^{\circ}$), ($85^{\circ}$, $75^{\circ}$), ($80^{\circ}$, $70^{\circ}$), ..., ($-80^{\circ}$, $-90^{\circ}$);
    \item in each latitudinal bin, total daily CH area are computed;
    \item daily CH areas are averaged using sliding window of size 1/2 year (180 days);
    \item daily solar wind speed data are averaged using the sliding window of the same size 1/2 year (180 days); 
    \item for each latitudinal bin, rolling correlation coefficients are calculated using sliding windows of size 2 years (i.e., in each time window of size 2 years, the correlation coefficient is calculated);
    \item correlation coefficients obtained for each time window of size 2 years and latitudinal bin are plotted in time-latitude diagram.
\end{itemize}

We observe in Figure~\ref{corr_map} that there are localized time-latitude regions in which CH areas are highly (above 0.75) correlated with solar wind speed. Some of these regions are located at polar latitudes (above $50^{\circ}$). We assume that this correlation is spurious and emerged due to variations of the solar B angle. It is possible that local increasing of the solar wind speed can randomly coincide with increasing of B angle and the corresponding increasing of the visible (not physical) CH area in the polar region. One more argument to ignore high correlation at polar latitudes is that in each case when we find high correlation at polar latitudes, there are also regions at low latitudes with high correlation as well. We assume that CHs at the low-latitudinal regions are potential candidates for the sources of HSS.

\begin{figure}
\centering
\includegraphics[width=0.95\columnwidth]{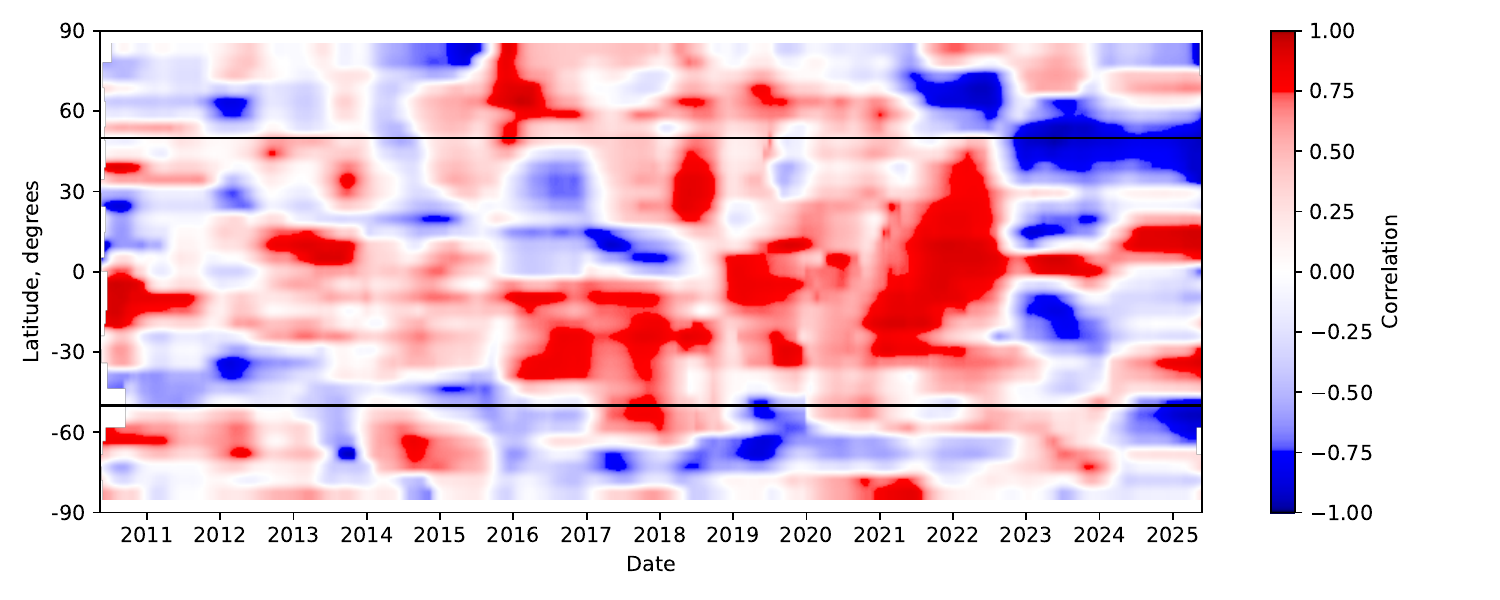}
\caption{Same as Figure~\ref{corr_map} but for the SPoCA-HEK catalog.}
\label{corr_map_spoca}
\end{figure}

For comparison, Figure~\ref{corr_map_spoca} shows a correlational time-latitude diagram based on the SPoCA-HEK catalog. We observe that visually it has a good agreement with Figure~\ref{corr_map}, but for quantitative comparison we refer to Figure~\ref{corr_tables}.

\section{Conclusions and Discussion}

We investigated to what extent CH data can explain smoothed variations of solar wind speed measured at 1 a.u. near Earth. In contrast to many previous studies, we were interested in comparison of smoothed (in time) data rather than instantaneous values. By smoothing data, we avoided complicated questions of the time delay between CH appearance and solar wind arrival and determination of the source CH for each HSS. As we demonstrated, smoothing also reduces contribution of other sources of high-speed solar wind streams such as ICME. 

We based our investigation on the CH catalog prepared at the Kislovodsk Mountain Astronomical Station for the period 2010--2025. For comparison, the SPoCA-HEK catalog was used. Both catalogs have a good agreement for polar CHs, but differ in the statistics of low-latitude CHs. 

The key result of our study is that we were able to construct a set of CH area indices that show variations similar to the time-averaged solar wind speed. These indices are based on the KMAS catalog and represent the area of CHs located within $30^{\circ}$ of the solar equator. Averaged over 1/2-year sliding windows, this index yields a corelation of 0.83 with the solar wind speed. Extending the size of sliding windows to 1 year, the correlation increases to 0.89, for 2-years sliding windows the correlation is 0.94. It is important that the latitudinal boundary of $30^{\circ}$ becomes lower than the typical boundary of $50^{\circ}$ separating polar CHs.

One could argue that the fact that low-latitude CHs are associated with variations of solar wind speed is well-known and widely used, e.g., in space weather prediction. The point is that this fact is usually demonstrated using a small number of HSS peaks for which the source CHs were identified, and parameters of those CHs were correlated with instantaneous solar wind speed. Although it reveals certain dependencies, but this approach is difficult to extend on larger time scales and conclude about overall correlation. In our study we obtained that the overall agreement can be demonstrated in a simple and convincing way using just CH areas at low latitudes.

\begin{figure}
\centering
\includegraphics[width=0.95\columnwidth]{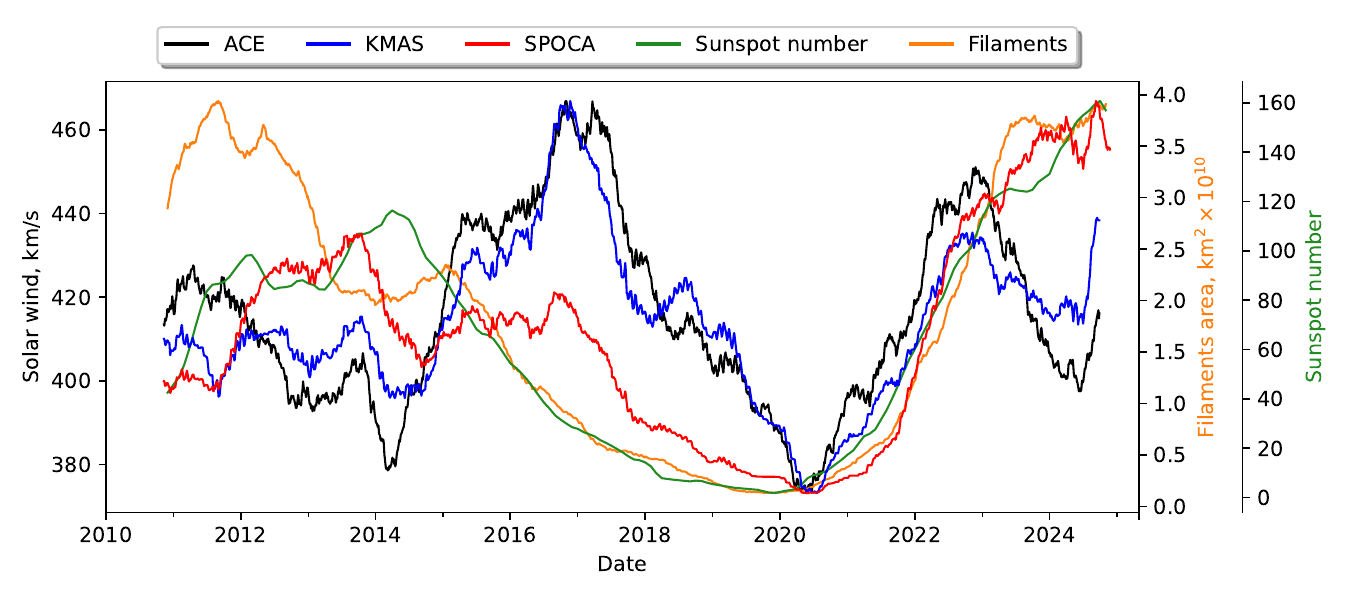}
\caption{Comparison of 1-year smoothed low-latitude ($|\theta| < 30^{\circ}$) coronal hole area indices, 1-year smoothed solar wind speed, 1-year smoothed total daily area of filaments, and 13-month smoothed monthly sunspot numbers.}
\label{ssn}
\end{figure}

At the same time, the ability to demonstrate the agreement between coronal hole areas at low latitudes and solar wind speed variations crucially depends on the particular catalog of CHs. The proposed indices of coronal hole areas show high correlation using the KMAS catalog, but, e.g., using the SPoCA-HEK catalog the correlation remains low even when large averaging windows are used. We see a possible reason in the amount of filaments identified as CHs in the SPoCA-HEK catalog \citep[see][for details]{Reiss_2024}. Partially, this situation can be observed in Figure~\ref{ssn} where we compare 1-year smoothed indices of low-latitude CH areas, 1-year smoothed total daily area of filaments observed at KMAS in H-alpha \citep{Tlatov2016}, and index of solar activity represented by the 13-month smoothed monthly sunspot numbers (SN) from the World Data Center SILSO, Royal Observatory of Belgium \citep{SILSO}. 

We observe in Figure~\ref{ssn} that the index of CH areas based on SPoCA-HEK catalog is more similar with filament areas and SN data than with solar wind speed, especially after 2020. Numerically, the correlation with filament areas is 0.78, with SN is 0.89, while with solar wind speed it is only 0.29. The correlation between filament areas and SN itself is 0.87 and the best match is observed during the declining phase of solar cycle 24 and the increasing phase of current cycle 25. 
Both CH indices and solar wind speed reach the minimum in mid-2020, which is about a half year later than the minimum of the  filament areas and the sunspot number index. After the minimum, all indices increase at least until the end of 2022, when the solar wind speed reaches the maximum and starts to decrease together with the KMAS index. In contrast, the SPoCA-HEK, SN, and filament indices continue to increase at least until the end of 2024. Also in cycle 24, the SPoCA-HEK index is maximal near the maximum of the solar cycle and filaments, rather than during the declining phase of the solar cycle, when the solar wind speed reaches its maximum.

We do not investigate other catalogs of CHs apart from KMAS and SPoCA-HEK, because our goal was not a comparative study, but to find at least one index that has good agreement with solar wind speed variations on large time scales. We do not exclude that using the SPoCA-HEK or other catalogs one can propose indices with the same or higher correlation, probably using more complex parameters of CHs \citep[e.g.,][demonstrated significance of the contrast of CH]{Obridko2009}. 

We also presented time-latitude diagrams (Figures~\ref{corr_map} and~\ref{corr_map_spoca}) of rolling correlation between CH areas and solar wind speed. These diagrams support the assumption that source CHs for solar wind variations can be found at low-latitudes. The time-elongated structures in these diagrams can be associated with recurrent solar wind streams produced by long-lived CHs. 

For completeness, the results presented for low-latitude CHs should be considered together with recent arguments revealing the connection of some small low-latitude CHs near solar maximum with the low-speed solar wind \citep[see, e.g., review][]{DAmicis2021, Bale2019, Wang2019}.

\begin{acks}
    EI acknowledges the support of RSF grant 21-72-20067.
\end{acks}

\section*{Data Availability}
    Coronal holes and filaments from the KMAS catalog are presented at the website \url{https://www.observethesun.com} and in the archive \url{http://en.solarstation.ru/archive}.
    Coronal holes from the SPoCA-HEK catalog are available at Heliophysics Event Knowledgebase \url{https://www.lmsal.com/hek}.
    ACE solar wind data are available at the OMNI
database \url{https://www.omniweb.gsfc.nasa.gov}.
    The catalog of near-Earth ICMEs is available at \url{https://srl.caltech.edu/ACE/ASC/DATA/level3/icmetable2.htm}.
    Sunspot number data are available at \url{https://www.sidc.be/SILSO/datafiles}.

\begin{conflict}
    The authors declare that they have no conflicts of interest.
\end{conflict}

\bibliographystyle{spr-mp-sola}
\bibliography{literature}

\begin{thebibliography}{29}
\ifx\bisbn     \undefined \def\bisbn  #1{ISBN #1}\fi
\ifx\binits    \undefined \def\binits#1{#1}\fi
\ifx\bauthor   \undefined \def\bauthor#1{#1}\fi
\ifx\batitle   \undefined \def\batitle#1{#1}\fi
\ifx\bjtitle   \undefined \def\bjtitle#1{\textit{#1}}\fi
\ifx\bvolume   \undefined \def\bvolume#1{\textbf{#1}}\fi
\ifx\byear     \undefined \def\byear#1{#1}\fi
\ifx\bissue    \undefined \def\bissue#1{#1}\fi
\ifx\bfpage    \undefined \def\bfpage#1{#1}\fi
\ifx\blpage    \undefined \def\blpage #1{#1}\fi
\ifx\burl      \undefined \def\burl#1{#1}\fi
\ifx\href      \undefined \def\href#1#2{#2}\fi
\ifx\betal     \undefined \def\betal{et al.}\fi
\ifx\bctitle   \undefined \def\bctitle#1{#1}\fi
\ifx\beditor   \undefined \def\beditor#1{#1}\fi
\ifx\bbtitle   \undefined \def\bbtitle#1{\textit{#1}}\fi
\ifx\bedition  \undefined \def\bedition#1{#1}\fi
\ifx\bseriesno \undefined \def\bseriesno#1{\textbf{#1}}\fi
\ifx\blocation \undefined \def\blocation#1{#1}\fi
\ifx\bsertitle \undefined \def\bsertitle#1{\textit{#1}}\fi
\ifx\bsnm      \undefined \def\bsnm#1{#1}\fi
\ifx\bsuffix   \undefined \def\bsuffix#1{#1}\fi
\ifx\bparticle \undefined \def\bparticle#1{#1}\fi
\ifx\barticle  \undefined \def\barticle#1{}\fi
\ifx\binstitute  \undefined \def\binstitute#1{#1}\fi
\ifx\bpublisher  \undefined \def\bpublisher#1{#1}\fi
\ifx\doiurl    \undefined \def\doiurl#1{\href{#1}{DOI}}\fi
\makeatletter
\def\safeHref#1#2#3{\in@{http}{#2}\ifin@\href{#2}{#3}\else\href{#1#2}{#3}\fi}
\makeatother
\ifx\adsurl    \undefined \def\adsurl#1{\safeHref{https://ui.adsabs.harvard.edu/abs/}{#1}{ADS}}\fi
\ifx\arxivurl  \undefined \def\arxivurl#1{\safeHref{http://arxiv.org/abs/}{#1}{arXiv}}\fi
\ifx\botherref \undefined \def\botherref#1{}\fi
\ifx\url       \undefined \def\url#1{#1}\fi
\ifx\bchapter  \undefined \def\bchapter#1{}\fi
\ifx\bbook     \undefined \def\bbook#1{}\fi
\ifx\bcomment  \undefined \def\bcomment#1{#1}\fi
\ifx\oauthor   \undefined \def\oauthor#1{#1}\fi
\ifx\citeauthoryear \undefined\def \citeauthoryear#1{#1}\fi
\def\endbibitem {}
\ifx\bconflocation  \undefined \def\bconflocation#1{#1} \fi

\bibitem[\protect\citeauthoryear{{Abramenko}, {Yurchyshyn}, and {Watanabe}}{2009}]{Abramenko}
\begin{barticle}
\bauthor{\bsnm{{Abramenko}}, \binits{V.}},
\bauthor{\bsnm{{Yurchyshyn}}, \binits{V.}},
\bauthor{\bsnm{{Watanabe}}, \binits{H.}}:
\byear{2009},
\batitle{{Parameters of the Magnetic Flux inside Coronal Holes}}.
\bjtitle{\solphys}
\bvolume{260},
\bfpage{43}.
\doiurl{https://doi.org/10.1007/s11207-009-9433-7}.
\adsurl{2009SoPh..260...43A}.
\end{barticle}
\endbibitem

\bibitem[\protect\citeauthoryear{{Bale} et~al.}{2019}]{Bale2019}
\begin{barticle}
\bauthor{\bsnm{{Bale}}, \binits{S.D.}},
\bauthor{\bsnm{{Badman}}, \binits{S.T.}},
\bauthor{\bsnm{{Bonnell}}, \binits{J.W.}},
\bauthor{\bsnm{{Bowen}}, \binits{T.A.}},
\bauthor{\bsnm{{Burgess}}, \binits{D.}},
\bauthor{\bsnm{{Case}}, \binits{A.W.}},
\bauthor{\bsnm{{Cattell}}, \binits{C.A.}},
\bauthor{\bsnm{{Chandran}}, \binits{B.D.G.}},
\bauthor{\bsnm{{Chaston}}, \binits{C.C.}},
\bauthor{\bsnm{{Chen}}, \binits{C.H.K.}},
\bauthor{\bsnm{{Drake}}, \binits{J.F.}},
\bauthor{\bsnm{{de Wit}}, \binits{T.D.}},
\bauthor{\bsnm{{Eastwood}}, \binits{J.P.}},
\bauthor{\bsnm{{Ergun}}, \binits{R.E.}},
\bauthor{\bsnm{{Farrell}}, \binits{W.M.}},
\bauthor{\bsnm{{Fong}}, \binits{C.}},
\bauthor{\bsnm{{Goetz}}, \binits{K.}},
\bauthor{\bsnm{{Goldstein}}, \binits{M.}},
\bauthor{\bsnm{{Goodrich}}, \binits{K.A.}},
\bauthor{\bsnm{{Harvey}}, \binits{P.R.}},
\bauthor{\bsnm{{Horbury}}, \binits{T.S.}},
\bauthor{\bsnm{{Howes}}, \binits{G.G.}},
\bauthor{\bsnm{{Kasper}}, \binits{J.C.}},
\bauthor{\bsnm{{Kellogg}}, \binits{P.J.}},
\bauthor{\bsnm{{Klimchuk}}, \binits{J.A.}},
\bauthor{\bsnm{{Korreck}}, \binits{K.E.}},
\bauthor{\bsnm{{Krasnoselskikh}}, \binits{V.V.}},
\bauthor{\bsnm{{Krucker}}, \binits{S.}},
\bauthor{\bsnm{{Laker}}, \binits{R.}},
\bauthor{\bsnm{{Larson}}, \binits{D.E.}},
\bauthor{\bsnm{{MacDowall}}, \binits{R.J.}},
\bauthor{\bsnm{{Maksimovic}}, \binits{M.}},
\bauthor{\bsnm{{Malaspina}}, \binits{D.M.}},
\bauthor{\bsnm{{Martinez-Oliveros}}, \binits{J.}},
\bauthor{\bsnm{{McComas}}, \binits{D.J.}},
\bauthor{\bsnm{{Meyer-Vernet}}, \binits{N.}},
\bauthor{\bsnm{{Moncuquet}}, \binits{M.}},
\bauthor{\bsnm{{Mozer}}, \binits{F.S.}},
\bauthor{\bsnm{{Phan}}, \binits{T.D.}},
\bauthor{\bsnm{{Pulupa}}, \binits{M.}},
\bauthor{\bsnm{{Raouafi}}, \binits{N.E.}},
\bauthor{\bsnm{{Salem}}, \binits{C.}},
\bauthor{\bsnm{{Stansby}}, \binits{D.}},
\bauthor{\bsnm{{Stevens}}, \binits{M.}},
\bauthor{\bsnm{{Szabo}}, \binits{A.}},
\bauthor{\bsnm{{Velli}}, \binits{M.}},
\bauthor{\bsnm{{Woolley}}, \binits{T.}},
\bauthor{\bsnm{{Wygant}}, \binits{J.R.}}:
\byear{2019},
\batitle{{Highly structured slow solar wind emerging from an equatorial coronal hole}}.
\bjtitle{\nat}
\bvolume{576},
\bfpage{237}.
\doiurl{https://doi.org/10.1038/s41586-019-1818-7}.
\adsurl{2019Natur.576..237B}.
\end{barticle}
\endbibitem

\bibitem[\protect\citeauthoryear{{Caplan}, {Downs}, and {Linker}}{2016}]{Caplan}
\begin{barticle}
\bauthor{\bsnm{{Caplan}}, \binits{R.M.}},
\bauthor{\bsnm{{Downs}}, \binits{C.}},
\bauthor{\bsnm{{Linker}}, \binits{J.A.}}:
\byear{2016},
\batitle{{Synchronic Coronal Hole Mapping Using Multi-instrument EUV Images: Data Preparation and Detection Method}}.
\bjtitle{\apj}
\bvolume{823},
\bfpage{53}.
\doiurl{https://doi.org/10.3847/0004-637X/823/1/53}.
\adsurl{2016ApJ...823...53C}.
\end{barticle}
\endbibitem

\bibitem[\protect\citeauthoryear{{Clette} and {Lefèvre}}{2015}]{SILSO}
\begin{botherref}
\oauthor{\bsnm{{Clette}}, \binits{F.}},
\oauthor{\bsnm{{Lefèvre}}, \binits{L.}}:
2015,
\textit{SILSO Sunspot Number V2.0},
https://doi.org/10.24414/qnza-ac80.
Published by WDC SILSO - Royal Observatory of Belgium (ROB).
\doiurl{https://doi.org/10.24414/qnza-ac80}.
\end{botherref}
\endbibitem

\bibitem[\protect\citeauthoryear{{Cranmer}}{2009}]{Cranmer}
\begin{barticle}
\bauthor{\bsnm{{Cranmer}}, \binits{S.R.}}:
\byear{2009},
\batitle{{Coronal Holes}}.
\bjtitle{Living Reviews in Solar Physics}
\bvolume{6},
\bfpage{3}.
\doiurl{https://doi.org/10.12942/lrsp-2009-3}.
\adsurl{2009LRSP....6....3C}.
\end{barticle}
\endbibitem

\bibitem[\protect\citeauthoryear{{D'Amicis} et~al.}{2021}]{DAmicis2021}
\begin{barticle}
\bauthor{\bsnm{{D'Amicis}}, \binits{R.}},
\bauthor{\bsnm{{Perrone}}, \binits{D.}},
\bauthor{\bsnm{{Bruno}}, \binits{R.}},
\bauthor{\bsnm{{Velli}}, \binits{M.}}:
\byear{2021},
\batitle{{On Alfv{\'e}nic Slow Wind: A Journey From the Earth Back to the Sun}}.
\bjtitle{Journal of Geophysical Research (Space Physics)}
\bvolume{126},
\bfpage{e28996}.
\doiurl{https://doi.org/10.1029/2020JA028996}.
\adsurl{2021JGRA..12628996D}.
\end{barticle}
\endbibitem

\bibitem[\protect\citeauthoryear{{Harvey} and {Recely}}{2002}]{Harvey}
\begin{barticle}
\bauthor{\bsnm{{Harvey}}, \binits{K.L.}},
\bauthor{\bsnm{{Recely}}, \binits{F.}}:
\byear{2002},
\batitle{{Polar Coronal Holes During Cycles 22 and 23}}.
\bjtitle{\solphys}
\bvolume{211},
\bfpage{31}.
\doiurl{https://doi.org/10.1023/A:1022469023581}.
\adsurl{2002SoPh..211...31H}.
\end{barticle}
\endbibitem

\bibitem[\protect\citeauthoryear{{Heinemann} et~al.}{2018}]{2018Heinemann}
\begin{barticle}
\bauthor{\bsnm{{Heinemann}}, \binits{S.G.}},
\bauthor{\bsnm{{Temmer}}, \binits{M.}},
\bauthor{\bsnm{{Hofmeister}}, \binits{S.J.}},
\bauthor{\bsnm{{Veronig}}, \binits{A.M.}},
\bauthor{\bsnm{{Vennerstr{\o}m}}, \binits{S.}}:
\byear{2018},
\batitle{{Three-phase Evolution of a Coronal Hole. I. 360{\textdegree} Remote Sensing and In Situ Observations}}.
\bjtitle{\apj}
\bvolume{861},
\bfpage{151}.
\doiurl{https://doi.org/10.3847/1538-4357/aac897}.
\adsurl{2018ApJ...861..151H}.
\end{barticle}
\endbibitem

\bibitem[\protect\citeauthoryear{{Heinemann} et~al.}{2020}]{2020Heinemann}
\begin{barticle}
\bauthor{\bsnm{{Heinemann}}, \binits{S.G.}},
\bauthor{\bsnm{{Jer{\v{c}}i{\'c}}}, \binits{V.}},
\bauthor{\bsnm{{Temmer}}, \binits{M.}},
\bauthor{\bsnm{{Hofmeister}}, \binits{S.J.}},
\bauthor{\bsnm{{Dumbovi{\'c}}}, \binits{M.}},
\bauthor{\bsnm{{Vennerstrom}}, \binits{S.}},
\bauthor{\bsnm{{Verbanac}}, \binits{G.}},
\bauthor{\bsnm{{Veronig}}, \binits{A.M.}}:
\byear{2020},
\batitle{{A statistical study of the long-term evolution of coronal hole properties as observed by SDO}}.
\bjtitle{\aap}
\bvolume{638},
\bfpage{A68}.
\doiurl{https://doi.org/10.1051/0004-6361/202037613}.
\adsurl{2020A&A...638A..68H}.
\end{barticle}
\endbibitem

\bibitem[\protect\citeauthoryear{Hess~Webber et~al.}{2014}]{HessWebber2014}
\begin{barticle}
\bauthor{\bsnm{Hess~Webber}, \binits{S.A.}},
\bauthor{\bsnm{Karna}, \binits{N.}},
\bauthor{\bsnm{Pesnell}, \binits{W.D.}},
\bauthor{\bsnm{Kirk}, \binits{M.S.}}:
\byear{2014},
\batitle{Areas of Polar Coronal Holes from 1996 Through 2010}.
\bjtitle{Solar Physics}
\bvolume{289},
\bfpage{4047–4067}.
\doiurl{https://doi.org/10.1007/s11207-014-0564-0}.
\burl{http://dx.doi.org/10.1007/s11207-014-0564-0}.
\end{barticle}
\endbibitem

\bibitem[\protect\citeauthoryear{{Hofmeister} et~al.}{2018}]{2018Hofmeister}
\begin{barticle}
\bauthor{\bsnm{{Hofmeister}}, \binits{S.J.}},
\bauthor{\bsnm{{Veronig}}, \binits{A.}},
\bauthor{\bsnm{{Temmer}}, \binits{M.}},
\bauthor{\bsnm{{Vennerstrom}}, \binits{S.}},
\bauthor{\bsnm{{Heber}}, \binits{B.}},
\bauthor{\bsnm{{Vr{\v{s}}nak}}, \binits{B.}}:
\byear{2018},
\batitle{{The Dependence of the Peak Velocity of High-Speed Solar Wind Streams as Measured in the Ecliptic by ACE and the STEREO satellites on the Area and Co-latitude of Their Solar Source Coronal Holes}}.
\bjtitle{Journal of Geophysical Research (Space Physics)}
\bvolume{123},
\bfpage{1738}.
\doiurl{https://doi.org/10.1002/2017JA024586}.
\adsurl{2018JGRA..123.1738H}.
\end{barticle}
\endbibitem

\bibitem[\protect\citeauthoryear{{Karachik} and {Pevtsov}}{2011}]{Karachik}
\begin{barticle}
\bauthor{\bsnm{{Karachik}}, \binits{N.V.}},
\bauthor{\bsnm{{Pevtsov}}, \binits{A.A.}}:
\byear{2011},
\batitle{{Solar Wind and Coronal Bright Points inside Coronal Holes}}.
\bjtitle{\apj}
\bvolume{735},
\bfpage{47}.
\doiurl{https://doi.org/10.1088/0004-637X/735/1/47}.
\adsurl{2011ApJ...735...47K}.
\end{barticle}
\endbibitem

\bibitem[\protect\citeauthoryear{{Krieger}, {Timothy}, and {Roelof}}{1973}]{Krieger}
\begin{barticle}
\bauthor{\bsnm{{Krieger}}, \binits{A.S.}},
\bauthor{\bsnm{{Timothy}}, \binits{A.F.}},
\bauthor{\bsnm{{Roelof}}, \binits{E.C.}}:
\byear{1973},
\batitle{{A Coronal Hole and Its Identification as the Source of a High Velocity Solar Wind Stream}}.
\bjtitle{\solphys}
\bvolume{29},
\bfpage{505}.
\doiurl{https://doi.org/10.1007/BF00150828}.
\adsurl{1973SoPh...29..505K}.
\end{barticle}
\endbibitem

\bibitem[\protect\citeauthoryear{Lemen et~al.}{2012}]{Lemen2012}
\begin{barticle}
\bauthor{\bsnm{Lemen}, \binits{J.R.}},
\bauthor{\bsnm{Title}, \binits{A.M.}},
\bauthor{\bsnm{Akin}, \binits{D.J.}},
\bauthor{\bsnm{Boerner}, \binits{P.F.}},
\bauthor{\bsnm{Chou}, \binits{C.}},
\bauthor{\bsnm{Drake}, \binits{J.F.}},
\bauthor{\bsnm{Duncan}, \binits{D.W.}},
\bauthor{\bsnm{Edwards}, \binits{C.G.}},
\bauthor{\bsnm{Friedlaender}, \binits{F.M.}},
\bauthor{\bsnm{Heyman}, \binits{G.F.}},
\bauthor{\bsnm{Hurlburt}, \binits{N.E.}},
\bauthor{\bsnm{Katz}, \binits{N.L.}},
\bauthor{\bsnm{Kushner}, \binits{G.D.}},
\bauthor{\bsnm{Levay}, \binits{M.}},
\bauthor{\bsnm{Lindgren}, \binits{R.W.}},
\bauthor{\bsnm{Mathur}, \binits{D.P.}},
\bauthor{\bsnm{McFeaters}, \binits{E.L.}},
\bauthor{\bsnm{Mitchell}, \binits{S.}},
\bauthor{\bsnm{Rehse}, \binits{R.A.}},
\bauthor{\bsnm{Schrijver}, \binits{C.J.}},
\bauthor{\bsnm{Springer}, \binits{L.A.}},
\bauthor{\bsnm{Stern}, \binits{R.A.}},
\bauthor{\bsnm{Tarbell}, \binits{T.D.}},
\bauthor{\bsnm{Wuelser}, \binits{J.-P.}},
\bauthor{\bsnm{Wolfson}, \binits{C.J.}},
\bauthor{\bsnm{Yanari}, \binits{C.}},
\bauthor{\bsnm{Bookbinder}, \binits{J.A.}},
\bauthor{\bsnm{Cheimets}, \binits{P.N.}},
\bauthor{\bsnm{Caldwell}, \binits{D.}},
\bauthor{\bsnm{Deluca}, \binits{E.E.}},
\bauthor{\bsnm{Gates}, \binits{R.}},
\bauthor{\bsnm{Golub}, \binits{L.}},
\bauthor{\bsnm{Park}, \binits{S.}},
\bauthor{\bsnm{Podgorski}, \binits{W.A.}},
\bauthor{\bsnm{Bush}, \binits{R.I.}},
\bauthor{\bsnm{Scherrer}, \binits{P.H.}},
\bauthor{\bsnm{Gummin}, \binits{M.A.}},
\bauthor{\bsnm{Smith}, \binits{P.}},
\bauthor{\bsnm{Auker}, \binits{G.}},
\bauthor{\bsnm{Jerram}, \binits{P.}},
\bauthor{\bsnm{Pool}, \binits{P.}},
\bauthor{\bsnm{Soufli}, \binits{R.}},
\bauthor{\bsnm{Windt}, \binits{D.L.}},
\bauthor{\bsnm{Beardsley}, \binits{S.}},
\bauthor{\bsnm{Clapp}, \binits{M.}},
\bauthor{\bsnm{Lang}, \binits{J.}},
\bauthor{\bsnm{Waltham}, \binits{N.}}:
\byear{2012},
\batitle{The Atmospheric Imaging Assembly (AIA) on the Solar Dynamics Observatory (SDO)}.
\bjtitle{Solar Physics}
\bvolume{275},
\bfpage{17}.
\doiurl{https://doi.org/10.1007/s11207-011-9776-8}.
\burl{https://doi.org/10.1007/s11207-011-9776-8}.
\end{barticle}
\endbibitem

\bibitem[\protect\citeauthoryear{{Nolte} et~al.}{1976}]{Nolte}
\begin{barticle}
\bauthor{\bsnm{{Nolte}}, \binits{J.T.}},
\bauthor{\bsnm{{Krieger}}, \binits{A.S.}},
\bauthor{\bsnm{{Timothy}}, \binits{A.F.}},
\bauthor{\bsnm{{Gold}}, \binits{R.E.}},
\bauthor{\bsnm{{Roelof}}, \binits{E.C.}},
\bauthor{\bsnm{{Vaiana}}, \binits{G.}},
\bauthor{\bsnm{{Lazarus}}, \binits{A.J.}},
\bauthor{\bsnm{{Sullivan}}, \binits{J.D.}},
\bauthor{\bsnm{{McIntosh}}, \binits{P.S.}}:
\byear{1976},
\batitle{{Coronal holes as sources of solar wind.}}
\bjtitle{\solphys}
\bvolume{46},
\bfpage{303}.
\doiurl{https://doi.org/10.1007/BF00149859}.
\adsurl{1976SoPh...46..303N}.
\end{barticle}
\endbibitem

\bibitem[\protect\citeauthoryear{{Obridko} et~al.}{2009}]{Obridko2009}
\begin{barticle}
\bauthor{\bsnm{{Obridko}}, \binits{V.N.}},
\bauthor{\bsnm{{Shelting}}, \binits{B.D.}},
\bauthor{\bsnm{{Livshits}}, \binits{I.M.}},
\bauthor{\bsnm{{Asgarov}}, \binits{A.B.}}:
\byear{2009},
\batitle{{Contrast of Coronal Holes and Parameters of Associated Solar Wind Streams}}.
\bjtitle{\solphys}
\bvolume{260},
\bfpage{191}.
\doiurl{https://doi.org/10.1007/s11207-009-9435-5}.
\adsurl{2009SoPh..260..191O}.
\end{barticle}
\endbibitem

\bibitem[\protect\citeauthoryear{Reiss et~al.}{2021}]{Reiss_2021}
\begin{barticle}
\bauthor{\bsnm{Reiss}, \binits{M.A.}},
\bauthor{\bsnm{Muglach}, \binits{K.}},
\bauthor{\bsnm{Möstl}, \binits{C.}},
\bauthor{\bsnm{Arge}, \binits{C.N.}},
\bauthor{\bsnm{Bailey}, \binits{R.}},
\bauthor{\bsnm{Delouille}, \binits{V.}},
\bauthor{\bsnm{Garton}, \binits{T.M.}},
\bauthor{\bsnm{Hamada}, \binits{A.}},
\bauthor{\bsnm{Hofmeister}, \binits{S.}},
\bauthor{\bsnm{Illarionov}, \binits{E.}},
\bauthor{\bsnm{Jarolim}, \binits{R.}},
\bauthor{\bsnm{Kirk}, \binits{M.S.F.}},
\bauthor{\bsnm{Kosovichev}, \binits{A.}},
\bauthor{\bsnm{Krista}, \binits{L.}},
\bauthor{\bsnm{Lee}, \binits{S.}},
\bauthor{\bsnm{Lowder}, \binits{C.}},
\bauthor{\bsnm{MacNeice}, \binits{P.J.}},
\bauthor{\bsnm{Veronig}, \binits{A.}},
\bauthor{\bsnm{Team}, \binits{C.I.C.H.B.W.}}:
\byear{2021},
\batitle{The Observational Uncertainty of Coronal Hole Boundaries in Automated Detection Schemes}.
\bjtitle{The Astrophysical Journal}
\bvolume{913},
\bfpage{28}.
\doiurl{https://doi.org/10.3847/1538-4357/abf2c8}.
\burl{https://dx.doi.org/10.3847/1538-4357/abf2c8}.
\end{barticle}
\endbibitem

\bibitem[\protect\citeauthoryear{Reiss et~al.}{2024}]{Reiss_2024}
\begin{barticle}
\bauthor{\bsnm{Reiss}, \binits{M.A.}},
\bauthor{\bsnm{Muglach}, \binits{K.}},
\bauthor{\bsnm{Mason}, \binits{E.}},
\bauthor{\bsnm{Davies}, \binits{E.E.}},
\bauthor{\bsnm{Chakraborty}, \binits{S.}},
\bauthor{\bsnm{Delouille}, \binits{V.}},
\bauthor{\bsnm{Downs}, \binits{C.}},
\bauthor{\bsnm{Garton}, \binits{T.M.}},
\bauthor{\bsnm{Grajeda}, \binits{J.A.}},
\bauthor{\bsnm{Hamada}, \binits{A.}},
\bauthor{\bsnm{Heinemann}, \binits{S.G.}},
\bauthor{\bsnm{Hofmeister}, \binits{S.}},
\bauthor{\bsnm{Illarionov}, \binits{E.}},
\bauthor{\bsnm{Jarolim}, \binits{R.}},
\bauthor{\bsnm{Krista}, \binits{L.}},
\bauthor{\bsnm{Lowder}, \binits{C.}},
\bauthor{\bsnm{Verwichte}, \binits{E.}},
\bauthor{\bsnm{Arge}, \binits{C.N.}},
\bauthor{\bsnm{Boucheron}, \binits{L.E.}},
\bauthor{\bsnm{Foullon}, \binits{C.}},
\bauthor{\bsnm{Kirk}, \binits{M.S.}},
\bauthor{\bsnm{Kosovichev}, \binits{A.}},
\bauthor{\bsnm{Leisner}, \binits{A.}},
\bauthor{\bsnm{Möstl}, \binits{C.}},
\bauthor{\bsnm{Turtle}, \binits{J.}},
\bauthor{\bsnm{Veronig}, \binits{A.}}:
\byear{2024},
\batitle{A Community Data Set for Comparing Automated Coronal Hole Detection Schemes}.
\bjtitle{The Astrophysical Journal Supplement Series}
\bvolume{271},
\bfpage{6}.
\doiurl{https://doi.org/10.3847/1538-4365/ad1408}.
\burl{https://dx.doi.org/10.3847/1538-4365/ad1408}.
\end{barticle}
\endbibitem

\bibitem[\protect\citeauthoryear{{Richardson} and {Cane}}{2010}]{Richardson}
\begin{barticle}
\bauthor{\bsnm{{Richardson}}, \binits{I.G.}},
\bauthor{\bsnm{{Cane}}, \binits{H.V.}}:
\byear{2010},
\batitle{{Near-Earth Interplanetary Coronal Mass Ejections During Solar Cycle 23 (1996 - 2009): Catalog and Summary of Properties}}.
\bjtitle{\solphys}
\bvolume{264},
\bfpage{189}.
\doiurl{https://doi.org/10.1007/s11207-010-9568-6}.
\adsurl{2010SoPh..264..189R}.
\end{barticle}
\endbibitem

\bibitem[\protect\citeauthoryear{{Rotter} et~al.}{2012}]{Rotter}
\begin{barticle}
\bauthor{\bsnm{{Rotter}}, \binits{T.}},
\bauthor{\bsnm{{Veronig}}, \binits{A.M.}},
\bauthor{\bsnm{{Temmer}}, \binits{M.}},
\bauthor{\bsnm{{Vr{\v{s}}nak}}, \binits{B.}}:
\byear{2012},
\batitle{{Relation Between Coronal Hole Areas on the Sun and the Solar Wind Parameters at 1 AU}}.
\bjtitle{\solphys}
\bvolume{281},
\bfpage{793}.
\doiurl{https://doi.org/10.1007/s11207-012-0101-y}.
\adsurl{2012SoPh..281..793R}.
\end{barticle}
\endbibitem

\bibitem[\protect\citeauthoryear{{Samara} et~al.}{2022}]{Evangelia}
\begin{barticle}
\bauthor{\bsnm{{Samara}}, \binits{E.}},
\bauthor{\bsnm{{Magdaleni{\'c}}}, \binits{J.}},
\bauthor{\bsnm{{Rodriguez}}, \binits{L.}},
\bauthor{\bsnm{{Heinemann}}, \binits{S.G.}},
\bauthor{\bsnm{{Georgoulis}}, \binits{M.K.}},
\bauthor{\bsnm{{Hofmeister}}, \binits{S.J.}},
\bauthor{\bsnm{{Poedts}}, \binits{S.}}:
\byear{2022},
\batitle{{Influence of coronal hole morphology on the solar wind speed at Earth}}.
\bjtitle{\aap}
\bvolume{662},
\bfpage{A68}.
\doiurl{https://doi.org/10.1051/0004-6361/202142793}.
\adsurl{2022A&A...662A..68S}.
\end{barticle}
\endbibitem

\bibitem[\protect\citeauthoryear{{Sheeley}, {Harvey}, and {Feldman}}{1976}]{Sheeley}
\begin{barticle}
\bauthor{\bsnm{{Sheeley}}, \binits{J.} \bsuffix{N.~R.}},
\bauthor{\bsnm{{Harvey}}, \binits{J.W.}},
\bauthor{\bsnm{{Feldman}}, \binits{W.C.}}:
\byear{1976},
\batitle{{Coronal holes, solar wind streams, and recurrent geomagnetic disturbances: 1973 1976}}.
\bjtitle{\solphys}
\bvolume{49},
\bfpage{271}.
\doiurl{https://doi.org/10.1007/BF00162451}.
\adsurl{1976SoPh...49..271S}.
\end{barticle}
\endbibitem

\bibitem[\protect\citeauthoryear{Stone et~al.}{1998}]{Stone1998}
\begin{barticle}
\bauthor{\bsnm{Stone}, \binits{E.C.}},
\bauthor{\bsnm{Frandsen}, \binits{A.M.}},
\bauthor{\bsnm{Mewaldt}, \binits{R.A.}},
\bauthor{\bsnm{Christian}, \binits{E.R.}},
\bauthor{\bsnm{Margolies}, \binits{D.}},
\bauthor{\bsnm{Ormes}, \binits{J.F.}},
\bauthor{\bsnm{Snow}, \binits{F.}}:
\byear{1998},
\batitle{The Advanced Composition Explorer}.
\bjtitle{Space Science Reviews}
\bvolume{86},
\bfpage{1}.
\doiurl{https://doi.org/10.1023/A:1005082526237}.
\burl{https://doi.org/10.1023/A:1005082526237}.
\end{barticle}
\endbibitem

\bibitem[\protect\citeauthoryear{{Tlatov}, {Kuzanyan}, and {Vasil'yeva}}{2016}]{Tlatov2016}
\begin{barticle}
\bauthor{\bsnm{{Tlatov}}, \binits{A.G.}},
\bauthor{\bsnm{{Kuzanyan}}, \binits{K.M.}},
\bauthor{\bsnm{{Vasil'yeva}}, \binits{V.V.}}:
\byear{2016},
\batitle{{Tilt Angles of Solar Filaments over the Period of 1919 - 2014}}.
\bjtitle{\solphys}
\bvolume{291},
\bfpage{1115}.
\doiurl{https://doi.org/10.1007/s11207-016-0880-7}.
\adsurl{2016SoPh..291.1115T}.
\end{barticle}
\endbibitem

\bibitem[\protect\citeauthoryear{{Tlatov}, {Tavastsherna}, and {Vasil'eva}}{2014}]{Tavastsherna}
\begin{barticle}
\bauthor{\bsnm{{Tlatov}}, \binits{A.}},
\bauthor{\bsnm{{Tavastsherna}}, \binits{K.}},
\bauthor{\bsnm{{Vasil'eva}}, \binits{V.}}:
\byear{2014},
\batitle{{Coronal Holes in Solar Cycles 21 to 23}}.
\bjtitle{\solphys}
\bvolume{289},
\bfpage{1349}.
\doiurl{https://doi.org/10.1007/s11207-013-0387-4}.
\adsurl{2014SoPh..289.1349T}.
\end{barticle}
\endbibitem

\bibitem[\protect\citeauthoryear{{Tousey}, {Sandlin}, and {Purcell}}{1968}]{Tousey}
\begin{bchapter}
\bauthor{\bsnm{{Tousey}}, \binits{R.}},
\bauthor{\bsnm{{Sandlin}}, \binits{G.D.}},
\bauthor{\bsnm{{Purcell}}, \binits{J.D.}}:
\byear{1968},
\bctitle{{On Some Aspects of XUV Spectroheliograms}}.
In: \beditor{\bsnm{{Kiepenheuer}}, \binits{K.O.}} (ed.)
\bbtitle{Structure and Development of Solar Active Regions}
\bseriesno{35},
\bfpage{411}.
\adsurl{1968IAUS...35..411T}.
\end{bchapter}
\endbibitem

\bibitem[\protect\citeauthoryear{{Verbeeck, C.} et~al.}{2014}]{SpoCA2014}
\begin{barticle}
\bauthor{\bsnm{{Verbeeck, C.}}},
\bauthor{\bsnm{{Delouille, V.}}},
\bauthor{\bsnm{{Mampaey, B.}}},
\bauthor{\bsnm{{De Visscher, R.}}}:
\byear{2014},
\batitle{The SPoCA-suite: Software for extraction, characterization, and tracking of active regions and coronal holes on EUV images}.
\bjtitle{A\&A}
\bvolume{561},
\bfpage{A29}.
\doiurl{https://doi.org/10.1051/0004-6361/201321243}.
\burl{https://doi.org/10.1051/0004-6361/201321243}.
\end{barticle}
\endbibitem

\bibitem[\protect\citeauthoryear{{Vr{\v{s}}nak}, {Temmer}, and {Veronig}}{2007}]{vrvsnak2007coronal}
\begin{barticle}
\bauthor{\bsnm{{Vr{\v{s}}nak}}, \binits{B.}},
\bauthor{\bsnm{{Temmer}}, \binits{M.}},
\bauthor{\bsnm{{Veronig}}, \binits{A.M.}}:
\byear{2007},
\batitle{{Coronal Holes and Solar Wind High-Speed Streams: II. Forecasting the Geomagnetic Effects}}.
\bjtitle{\solphys}
\bvolume{240},
\bfpage{331}.
\doiurl{https://doi.org/10.1007/s11207-007-0311-x}.
\adsurl{2007SoPh..240..331V}.
\end{barticle}
\endbibitem

\bibitem[\protect\citeauthoryear{{Wang} and {Ko}}{2019}]{Wang2019}
\begin{barticle}
\bauthor{\bsnm{{Wang}}, \binits{Y.-M.}},
\bauthor{\bsnm{{Ko}}, \binits{Y.-K.}}:
\byear{2019},
\batitle{{Observations of Slow Solar Wind from Equatorial Coronal Holes}}.
\bjtitle{\apj}
\bvolume{880},
\bfpage{146}.
\doiurl{https://doi.org/10.3847/1538-4357/ab2add}.
\adsurl{2019ApJ...880..146W}.
\end{barticle}
\endbibitem

\end{thebibliography}

\end{document}